\documentclass[showpacs,preprintnumbers,amsmath,reprint,aps,prx,superscriptaddress]{revtex4-1}

\usepackage{bm}
\usepackage{graphicx}
\usepackage{color}
\usepackage{amsmath}
\usepackage{hyperref}
\usepackage{epstopdf}
\usepackage{upgreek}
\usepackage{mathptmx, textcomp}
\usepackage[latin1]{inputenc}
\usepackage[T1]{fontenc}
\usepackage{array, multirow}
\usepackage{booktabs}
\usepackage{longtable}
\usepackage[normalem]{ulem}
\usepackage{natbib}
\usepackage{amssymb}

\begin{document}

\title{Indium tin oxide films meet circular Rydberg atoms: prospects for novel quantum simulation schemes}

\author{Florian Meinert}
\affiliation{5. Physikalisches Institut, Universit\"{a}t Stuttgart, Pfaffenwaldring 57, 70569 Stuttgart, Germany}
\affiliation{Center for Integrated Quantum Science and Technology, IQST, Pfaffenwaldring 57, 70569 Stuttgart, Germany}
\author{Christian H\"olzl}
\affiliation{5. Physikalisches Institut, Universit\"{a}t Stuttgart, Pfaffenwaldring 57, 70569 Stuttgart, Germany}
\affiliation{Center for Integrated Quantum Science and Technology, IQST, Pfaffenwaldring 57, 70569 Stuttgart, Germany}
\author{Mehmet Ali Nebioglu}
\affiliation{1. Physikalisches Institut, Universit\"{a}t Stuttgart, Pfaffenwaldring 57, 70569 Stuttgart, Germany}
\author{Alessandro D'Arnese}
\affiliation{1. Physikalisches Institut, Universit\"{a}t Stuttgart, Pfaffenwaldring 57, 70569 Stuttgart, Germany}
\author{Philipp Karl}
\affiliation{1. Physikalisches Institut, Universit\"{a}t Stuttgart, Pfaffenwaldring 57, 70569 Stuttgart, Germany}
\author{Martin Dressel}
\affiliation{Center for Integrated Quantum Science and Technology, IQST, Pfaffenwaldring 57, 70569 Stuttgart, Germany}
\affiliation{1. Physikalisches Institut, Universit\"{a}t Stuttgart, Pfaffenwaldring 57, 70569 Stuttgart, Germany}
\author{Marc Scheffler}
\affiliation{1. Physikalisches Institut, Universit\"{a}t Stuttgart, Pfaffenwaldring 57, 70569 Stuttgart, Germany}
\date{\today}

\begin{abstract}
Long-lived circular Rydberg atoms are picking up increasing interest for boosting coherence times in Rydberg-based quantum simulation. We elaborate a novel approach to stabilize circular Rydberg states against spontaneous and blackbody-induced decay using a suppression capacitor made from indium tin oxide (ITO) thin films, which combine reflection of microwaves with transparency in the visible spectral range. To this end, we perform detailed characterization of such films using complementary spectroscopic methods at GHz and THz frequencies and identify conditions that allow for reaching circular-state lifetimes up to tens of milliseconds in a room-temperature environment. We discuss prospects of our findings in view of the quest for quantum simulations with high-$n$ circular Rydberg states at room temperature.
\end{abstract}

\maketitle 

\section{Introduction}

Ultracold Rydberg atoms have recently shown fascinating means for realizing a versatile platform for applications in quantum simulation and quantum information processing. Examples range from implementations of quantum spin models \cite{Labuhn2016,Bernien2017}, over studies of topological systems \cite{Leseleuc18}, to the demonstration of high-fidelity entangling gates \cite{Levine19,Graham19}. At the very heart of the achieved experimental control are methods for arranging individual atoms trapped in flexible arrays of optical tweezers in a bottom-up manner combined with strong interactions between pairs of Rydberg states \cite{Endres16,Barredo16,Barredo18}. An evident drawback of this appealing approach, however, is the Rydberg state decay, which sets fundamental limits on the coherence time of the quantum system. In this context, circular Rydberg states (i.e. Rydberg states with $|m|=n-1$, where $m$ and $n$ denote the orbital magnetic and principal quantum number) have recently been proposed as a possible candidate to circumvent this problem as they may allow for much longer lifetimes compared to the so far exploited low-angular momentum Rydberg levels (i.e. $S$-, $P$-, or $D$-orbitals) \cite{Nguyen18}.

For decades, circular Rydberg atoms have been extensively explored in atomic beam experiments \cite{Hulet83,Brecha93,Nussenzveig93}. Their individual interaction with microwave photons trapped in high-quality resonators allowed for fundamental studies of cavity quantum electrodynamics \cite{HarocheBook06}. More recently, superpositions of circular states were also instrumental for quantum sensing applications \cite{Dietsche19}. Experiments with ultracold and trapped circular Rydberg atoms are, however, comparatively rare \cite{Anderson13}. Efficient creation and optical trapping of long-lived circular states was demonstrated only very recently reporting lifetimes of several milliseconds in a $^4$He cryogenic environment \cite{Cortinas19,Cantat20}. In free space, radiative decay of the circular state is largely dominated by microwave transitions into neighboring $n$-manifolds. Yet, such long-wavelength decay paths for Rydberg states can be effectively suppressed by placing the atom inside a capacitor structure which prohibits the propagation of the corresponding microwave field mode \cite{Kleppner85,Hulet85}. Ultimately, a quantum simulator based on very long-lived circular Rydberg states should thus allow for optical trapping inside a microwave resonator, which poses stringent limitations for optical access and thus for trapping ensembles of atoms in defect-free tweezer arrays \cite{Nguyen18}.

In this paper, we propose a novel route to stabilize circular Rydberg states in a way that simultaneously allows for high numerical aperture optical access using optically transparent conductive thin films. Specifically, we explore the feasibility to realize a capacitor for suppressing high-$n$ circular Rydberg state decay with indium tin oxide (ITO) films (Fig.~\ref{Fig_decay_paths}(a)). Such a capacitor has to combine high reflectivity in the microwave domain with transparency for visible to short near-infrared wavelengths. While dc and optical properties of ITO films have been explored in great detail \cite{HartnagelBook1995,Tahar1998}, extended studies of their response at microwave frequencies are comparatively rare. We report on detailed millimeter-wave investigations of ITO thin films of different thickness and sheet resistance using a combination of complementary spectroscopy methods, which allow us to infer microwave properties of the films covering a frequency range from 100~MHz up to 1~THz. Our measurements demonstrate that ITO coatings are well suited to stabilize circular Rydberg states for tens of milliseconds even in a room temperature environment.

\section{Long-lived circular Rydberg states}

\subsection{Radiative decay of circular Rydberg states}

\begin{figure}[!ht]
	\includegraphics[width=0.495\columnwidth]{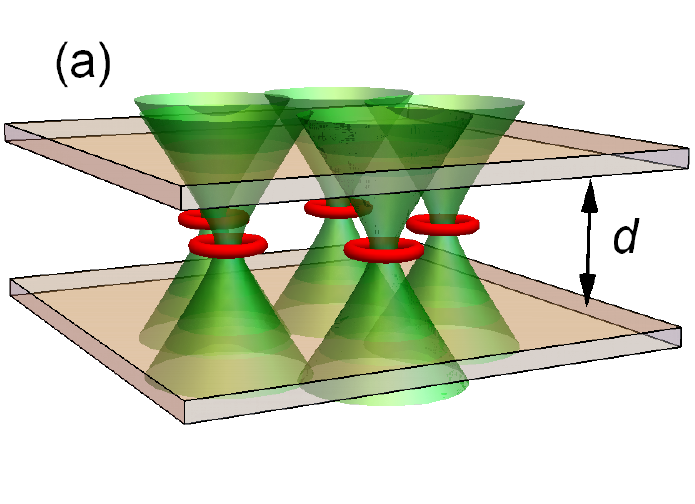}
	\includegraphics[width=0.495\columnwidth]{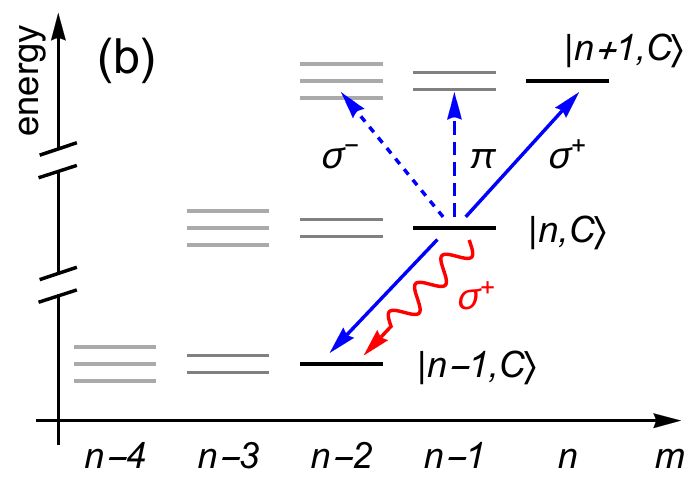}\\
	\includegraphics[width=0.495\columnwidth]{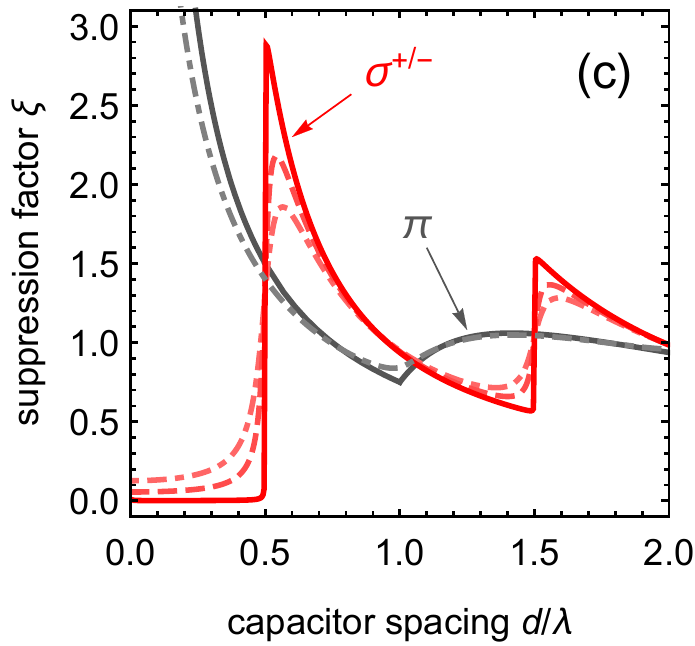}
	\includegraphics[width=0.495\columnwidth]{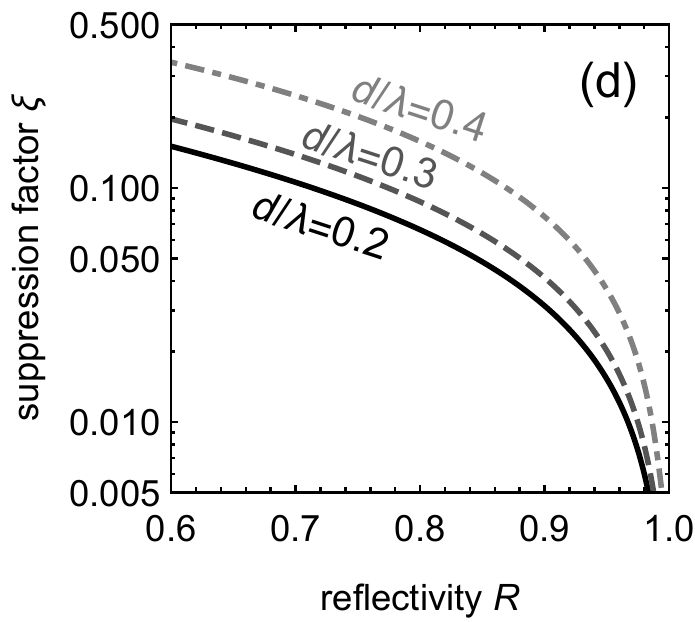}\\
	\caption{(a) Illustration of the proposed setup. A pair of ITO-coated glass plates separated by $d$ form a capacitor to stabilize the circular Rydberg state against radiative decay. While reflective at microwave frequencies, optical transparency provides access for trapping in optical tweezer arrays. (b) Atomic level scheme of the three largest angular momentum states for the Rydberg manifolds $n$ and $n\pm1$. Arrows denote open decay paths of the circular state $|n,C\rangle$ via spontaneous emission (red) and blackbody decay (blue). (c) Cavity-induced suppression factor $\xi$ for $\sigma^\pm$ (red) and $\pi$-transitions (gray) as a function of the capacitor spacing $d$. Solid lines depict $\xi$ for perfect reflection ($R=1$), while dashed (dot-dashed) lines show results for finite film reflectivity $R=0.8$ ($R=0.6$). (d) Suppression factor $\xi$ for $\sigma^\pm$-transitions as a function of $R$ for different $d$ as indicated.}
	\label{Fig_decay_paths}
\end{figure}

Starting point for our discussion is a brief summary of the unique properties of circular Rydberg states and the consequences for their radiative decay. For a given hydrogenic Rydberg manifold of high-$l$ states with principal quantum number $n$, the circular states $|n,C\rangle$ possess maximum angular momentum and magnetic quantum number, i.e. $l=|m|=n-1$ \cite{GallagherBook94}. In contrast to low-$l$ Rydberg levels, circular states are effectively protected from spontaneous emission at optical frequencies due to their large orbital angular momentum. Indeed, spontaneous decay can only happen via circularly polarized microwave radiation through a single dipole-allowed transition into the circular state of the next lower hydrogenic manifold $|n-1,C\rangle$ (see Fig.~\ref{Fig_decay_paths}(b)). Despite the large transition strength, the small mode density in the microwave domain causes much longer lifetimes compared to low-$l$ Rydberg states.

Evidently, this only holds for negligible blackbody-induced decay. Stimulated absorption and emission of $\sigma^+$-polarized  blackbody radiation at temperature $T$ drives the circular state into the adjacent levels $|n\pm1,C\rangle$. Furthermore, weaker transitions induced by the $\pi$- and $\sigma^-$-polarized components of the thermal radiation field couple to elliptical states (dashed and dotted arrows in Fig.~\ref{Fig_decay_paths}(b)). The total blackbody-reduced decay rate of the circular Rydberg level thus reads
\begin{equation}
\gamma = A_{|n,C\rangle,|n-1,C\rangle} + \sum\limits_{|f\rangle} \bar{n}(\omega,T) A_{|n,C\rangle,|f\rangle} \, .
\label{eq_gammablackbody}
\end{equation}
Here, $A_{|i\rangle,|f\rangle}$ denotes the Einstein $A$-coefficient between states $|i\rangle$ and $|f\rangle$, $\omega$ the corresponding transition frequency, and $\bar{n}$ the thermal average photon number. The sum runs over all final states $|f\rangle$ with non-zero dipole matrix element $\langle f| e\vec{r}|n,C\rangle$. Note that the latter decrease very rapidly for final states of more distant manifolds than $n+1$. In Fig.~\ref{Fig_calculation_lifetime}, the calculated lifetime 1$/\gamma$ for circular states ranging from $n=40$ to $n=120$ is shown for bare spontaneous decay ($T=0$, circles)  as well as for a situation at room temperature ($T=300 \, \rm{K}$, squares). The detrimental blackbody-induced reduction of the lifetime by up to three orders of magnitude ensues predominantly from the strong transitions $|n,C\rangle - |n\pm1,C\rangle$ between adjacent circular Rydberg states.

\subsection{Stabilizing circular Rydberg states in a cavity}

The above discussion suggests a severe obstacle in exploiting circular Rydberg states at room temperature to encode a long-lived qubit for quantum simulations. However, the challenge may be solved effectively by placing the atom in the center of a pair of conductive plane-parallel capacitor plates. As first predicted in Ref.~\cite{Kleppner85} and later observed experimentally \cite{Hulet85}, this setting allows for suppressing Rydberg state decay. The suppression is induced by suitably restricting the available electromagnetic field modes for atomic state decay. Specifically, a plane-parallel capacitor with spacing $d$ inhibits the propagation of modes with wavelengths $\lambda>2d$ and polarized parallel to the capacitor plates. For a circular Rydberg state $|n,C\rangle$ whose orbital plane is oriented parallel with the capacitor plates, this suppresses the dominant decay channels via circularly polarized photon modes when the microwave-photon wavelengths associated with the transitions $|n,C\rangle - |n\pm1,C\rangle$ fulfill the above condition. For example, the critical capacitor spacing associated with the transition wavelength between adjacent $n$-manifolds is $2.7$~mm for $n=50$ and $2.2$~cm for $n=100$, and the corresponding microwave transition frequencies $f=\omega/(2\pi)$ are 54.3~GHz and 6.7~GHz, respectively. Modes polarized perpendicular to the capacitor plane are generally not inhibited to propagate and induce the weaker $\pi$-transitions. Note that transitions via these modes may be even slightly enhanced by the cavity. Orientation of the circular state's orbital plane with respect to the cavity is readily achieved by applying a bias magnetic field of a few Gauss pointing perpendicular to the capacitor plates.

The suppression factor can be calculated in a classical framework considering an oscillating dipole in the vicinity of metallic surfaces. It is straightforward using the method of image charges to evaluate the case of the two infinitely extended plane-parallel capacitor plates with finite reflectivity $R$ (the field reflectivity is $r=\sqrt{R}$). One finds \cite{Haroche92,Hinds94}
\begin{eqnarray}
\xi^{\sigma^\pm} &=& 1 + 3 \operatorname{Im} \sum\limits_{n=1}^\infty (-r)^n \left(\frac{1}{\phi_n} + \frac{i}{\phi_n^2} -\frac{1}{\phi_n^3}\right) e^{i \phi_n} \\
\xi^{\pi} &=& 1 + 6 \operatorname{Im} \sum\limits_{n=1}^\infty r^n \left(- \frac{i}{\phi_n^2} + \frac{1}{\phi_n^3}\right) e^{i \phi_n}
\label{eq_suppressionfactor}
\end{eqnarray}
for $\sigma^\pm$- and $\pi$-transitions, respectively, and with $\phi_n = n 2 \pi d /\lambda$. The suppression factors for the different polarization modes and for various values of $R$ are depicted in Fig.~\ref{Fig_decay_paths}(c). For a perfectly reflecting capacitor, $\xi^{\sigma^\pm}$ is strictly zero for $d<\lambda/2$, and Rydberg-state decay via $\sigma^\pm$-transitions is fully suppressed. A finite reflectivity, however, causes imperfect suppression (see also Fig.~\ref{Fig_decay_paths}(d)) and partially opens the decay channel. Note also the aforementioned enhancement for $\pi$-transitions in the cavity.

\begin{figure}[!ht]
	\includegraphics[width=0.7\columnwidth]{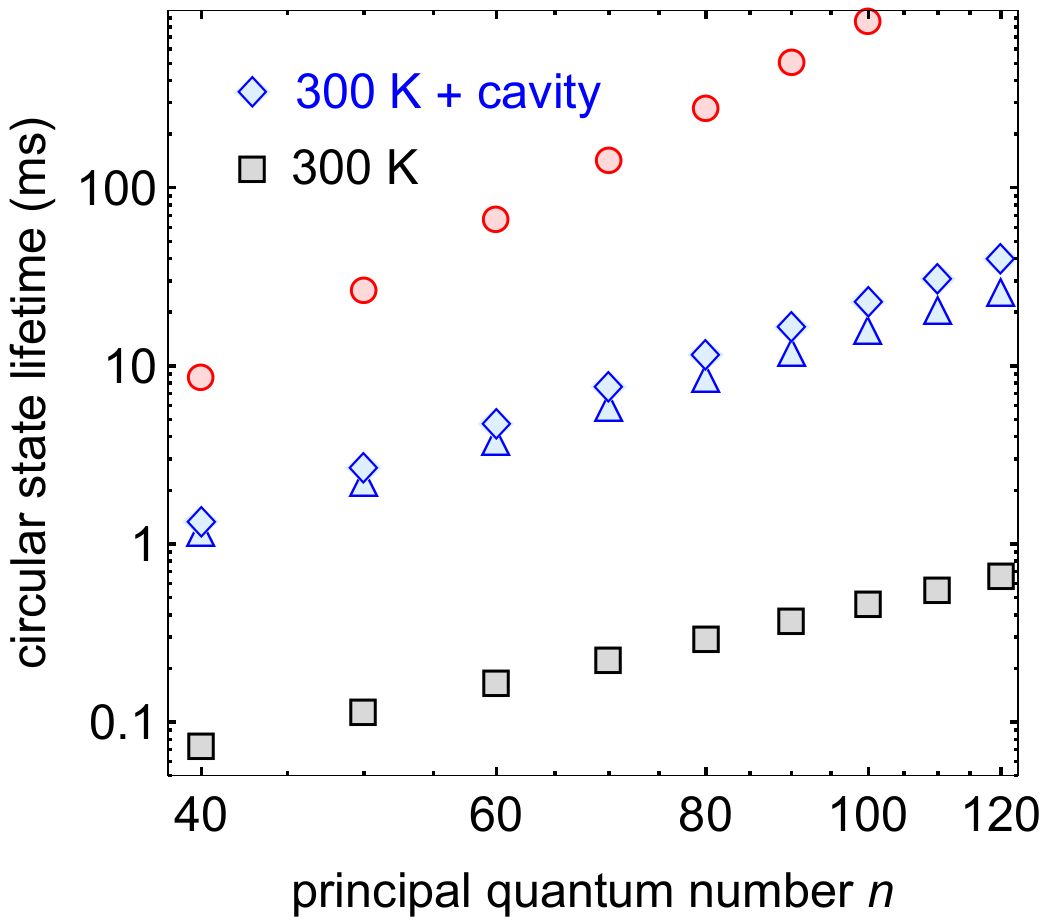}
	\caption{Calculated lifetime for circular Rydberg states $|n,C\rangle$ at room temperature ($T=300 \, \rm{K}$) in free space (squares), in an ideal plane-parallel cavity ($R=1$ and $d/\lambda=0.2$) (diamonds), and in a cavity with finite reflectivity ($R=0.97$ and $d/\lambda=0.2$) (triangles). The circles show the lifetime in free space due to spontaneous emission only, i.e. for $T=0 \, \rm{K}$.}
	\label{Fig_calculation_lifetime}
\end{figure}

We are now in a position to compute the circular-state lifetime in the capacitor by evaluating Eq.~\ref{eq_gammablackbody} with the cavity-modified Einstein $A$-coefficients, i.e. $\xi^{\sigma^\pm (\pi)} A_{|n,C\rangle,|f\rangle}$ for the respective $\sigma^{\pm}$($\pi$)-transitions \cite{Nguyen18,Mozley05}. The results at room temperature for an ideal cavity (diamonds) and for finite reflectivity (triangles) are shown in Fig.~\ref{Fig_calculation_lifetime}. Evidently, stabilizing the circular Rydberg state in a cavity is expected to allow for boosting the lifetime by up to two orders of magnitude at room temperature. Note that the value $d/\lambda=0.2$ in Fig.~\ref{Fig_calculation_lifetime} is chosen so that the capacitor also suppresses the $\sigma^+$-transition $|n,C\rangle - |n+2,l=n,m=n\rangle$, which would otherwise contribute to the overall decay with a rate comparable to the unsuppressed $\pi$-transition into the $n+1$ manifold.

\section{ITO thin films at \NoCaseChange{GHz} and \NoCaseChange{THz} frequencies}

In the following, we will explore whether the above conditions can be realized in an optically transparent capacitor, i.e.\ to which extent the capacitor plates can be at the same time highly reflective for GHz radiation and transparent for visible light. The strategy we pursue here is deposition of an appropriate conductive film on a glass substrate. The common high reflectivity of regular metals for visible light is due to the high density $N$ of conduction electrons, which sets the limiting frequency for metallic reflection, the plasma frequency, via $\omega_\textrm{p} = \sqrt{N e^2 / (\epsilon_0 \, m)}$ with $e$ the electron charge, $m$ their effective mass, and $\epsilon_0$ the vacuum permeability. Thus we seek a metal with $\omega_p/(2\pi)$ lower than $\sim 250$~THz, but $N$ should still be as high as possible to obtain high GHz reflectivity. These requirements are equivalent to those for transparent electrodes, where high transparency for visible light is to be combined with high dc conductivity \cite{Freeman2000,Gordon2000,Minami2005}. The classic solution for the latter application are ITO thin films. Here the exact constituent composition and preparation conditions allow material tuning to optimize the required material properties. While there is vast literature concerning the dc and optical properties of ITO films \cite{HartnagelBook1995,Tahar1998}, the THz and GHz properties have been addressed much less \cite{Takizawa1999,Bauer2002,Yang2013,Brown2015,Wang2015,Sheokand2017,Zhang2017,Lai2017}.

ITO is a doped semiconductor \cite{Hamberg1984} with typical density of mobile charge carriers so high that the low-energy optical properties can be modeled by the Drude theory of metals, which predicts a frequency-dependent optical conductivity $\sigma(\omega) = \sigma_\textrm{dc} / (1 - i \omega \tau)$ with dc conductivity $\sigma_\textrm{dc} = 1 /\rho_\textrm{dc}$ and relaxation time $\tau$. Due to the strong disorder in ITO, the Drude relaxation rate $\Gamma = 1/\tau$ can be expected in the 10~THz range \cite{Wang2015}. Therefore, the GHz response of ITO should be well within the Hagen-Rubens regime, and thus a rather straightforward connection between dc and GHz properties is expected \cite{HartnagelBook1995}. In the following, we show that this is indeed the case by performing detailed GHz and THz measurements on ITO thin films. This will allow for a robust determination of the GHz reflectivity of ITO films.

\subsection{Experimental methods}
Far-field reflectivity measurements at GHz frequencies on cm-sized material samples are challenging because of the large wavelength. Therefore, we combine different experimental approaches for a comprehensive Hagen-Rubens characterization of the ITO film. At THz frequencies, far-field optical spectroscopy is well established. Here we use two different techniques. First, frequency-domain spectroscopy (THz-FDS) using backward-wave oscillators (BWOs) as monochromatic, coherent, and frequency-tunable radiation sources allows for direct measurements of transmission and reflection coefficients (under normal incidence) \cite{Gorshunov2005,Pracht2013}. Using several BWOs we cover frequencies between 100~GHz and 500~GHz. Second, we perform transmission measurements using a THz time-domain spectrometer (THz-TDS) \cite{Hangyo2005}.

For the low-GHz spectral range, we employ reflection measurements in so-called Corbino geometry \cite{Scheffler2005}. Here, coaxial Au contacts on the ITO film define the actual Corbino disk, which is then directly pressed against a coaxial probe at the end of a coaxial microwave transmission line. A vector network analyzer (VNA) generates monochromatic GHz signals that travel in this line, are reflected by the sample, and then detected by the VNA. The measured quantity is the reflection coefficient $S_\mathrm{11}$, which is the ratio of the voltages of reflected and incoming microwave signal and which is not to be confused with the optical reflectivity in far-field configuration. For our case of thin metallic films, one can directly obtain the conductivity $\sigma$ from each measured $S_\mathrm{11}$ \cite{Scheffler2005}. All experiments are performed at room temperature.

\begin{table}[!ht]
\begin{tabular}{c||c|c|c|c|c|}
  & $t$ & $R_\square$ 		& $\rho_\textrm{dc}$			& $T_{780 \rm{nm}}$ & $T_{461 \rm{nm}}$ \\
sample & (nm) & ($\Omega$) & ($\mu\Omega$cm) & (\%) & (\%) \\ \hline
 A & 135 & 44 & 600 & 87 & 82 \\ \hline
 B & 135 & 31 & 420 & 81 & 84 \\ \hline
 C & 135 & 9.8 & 130 & 87 & 85 \\ \hline
 D & 2000 & 1.9 & 390 & 73 & 60
\end{tabular}
\caption{Properties of the four characterized ITO-coated glass samples labeled A-D. Film thickness $t$ (as indicated by the supplier) and sheet resistance $R_\square$ (measured in van der Pauw geometry, with an estimated error of a few percent). The dc resistivity $\rho_\textrm{dc}$ is calculated from sheet resistance and thickness. Values for the transmission in the visible domain is given for 780 nm ($T_{780 \rm{nm}}$) and 461 nm ($T_{641 \rm{nm}}$), representative for fluorescence detection of Rb and Sr atoms.}
\label{Tab_ITOsamples}
\end{table}

\subsection{\NoCaseChange{GHz} and \NoCaseChange{THz} characteristics of ITO thin films}

We have studied four different ITO thin films that were grown on 1~mm thick glass substrates \cite{EndnoteVacuLayer}. Relevant film parameters are listed in Table \ref{Tab_ITOsamples}. The original 25~mm x 25~mm samples were cut into smaller pieces. For the THz experiments, 11~mm x 11~mm sized samples are used. On these samples we also perform 4-point dc resistance measurements in van der Pauw geometry. For the GHz measurements, smaller samples (here approximately 3~mm x 4~mm) are required \cite{Scheffler2005}, with Corbino contacts made of 200~nm thick, thermally evaporated Au deposited through a shadow mask.

\subsubsection{THz reflection and transmission}

\begin{figure}[!ht]
	\includegraphics[width=\columnwidth]{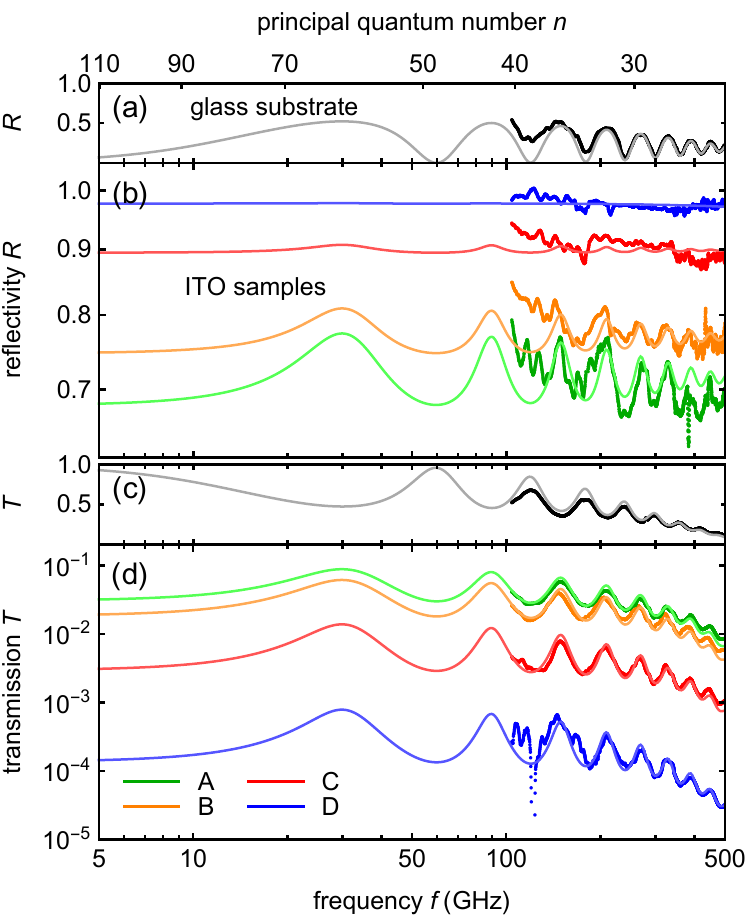}
	\caption{Reflectivity $R$ (b) and transmission $T$ (d) as a function of frequency $f$, obtained from frequency-domain spectroscopy, for the four ITO-coated glass samples A-D as indicated. For comparison, reflectivity and transmission of the bare uncoated glass substrate is shown in (a) and (c). The solid lines are computed spectra based on a Drude behavior and matched to the measurements (see text for details). The experimental data are smoothened using a running average. The upper abscissa indicates the principal quantum number $n$ of the transition $|n,C\rangle - |n-1,C\rangle$ with transition frequency $f$.}
	\label{Fig_reflection_transmission}
\end{figure}

In Fig.~\ref{Fig_reflection_transmission}, we present the THz transmission and reflection data obtained in the frequency domain together with theory curves that extend to the GHz regime of the Rydberg transitions between adjacent high-$n$ manifolds. While Figs.~\ref{Fig_reflection_transmission}(b) and (d) show reflection and transmission spectra, respectively, for the four investigated ITO-coated glass samples, Figs.~\ref{Fig_reflection_transmission}(a) and (c) show the corresponding data for a bare glass substrate. All spectra exhibit pronounced Fabry-Perot resonances due to standing waves of the coherent THz radiation in the dielectric glass substrate. The periodicity is governed by the glass thickness and dielectric constant. When compared to the spectra of the bare glass, the ITO-coated samples demonstrate the desired effect, namely a strongly enhanced reflection. At the same time it becomes clear that reflectivities beyond 0.95 are only obtained for the thickest ITO film (sample D, film thickness $t=2000$~nm). Measuring THz reflection close to unity is experimentally challenging as evident from some of the experimental data for sample D that are larger than one. Therefore, it is helpful to analyze the complementary THz transmission, where reference measurements are less error-prone.

To this end, the material parameters (dc conductivity $\sigma_{\rm{dc}}$ and relaxation rate $\Gamma$) entering the theory curves shown in Fig.~\ref{Fig_reflection_transmission} are tuned in order to yield a match with the measured transmission spectra. For the glass substrate curves, just a single strongly damped Lorentzian at very high frequency is assumed, with resulting low-frequency dielectric constant ($\epsilon_1 \approx 6.3$). These glass properties are then included in the model for the ITO spectra. For the ITO films, we assume simple Drude behavior with a relaxation rate $\Gamma/(2\pi)$ around 15~THz. Note that this value is much higher than the experimentally covered frequency range, and consequently, it should be considered only as a rough estimate. As the theory curves closely describe both, transmission and reflection spectra, the simple model assumptions are demonstrated to work rather well.

\subsubsection{Optical conductivity}

\begin{figure}[!ht]
\centering
	\includegraphics[width=\columnwidth]{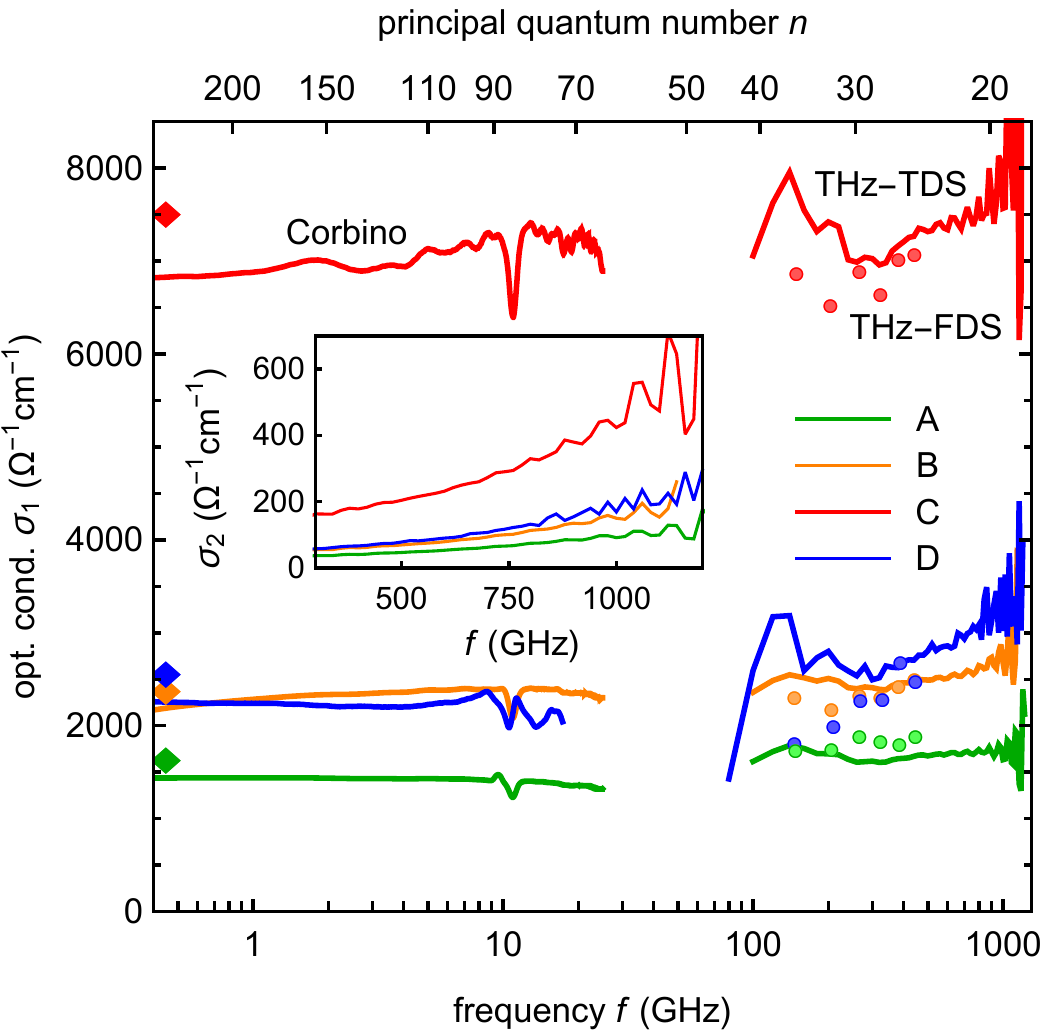}
	\caption{Optical conductivity $\sigma_1$ (real part) as a function of frequency $f$ for the four ITO-coated glass samples A-D as indicated. The datasets for $f<25 \, \rm{GHz}$ are obtained via Corbino spectroscopy. Filled circles indicate data determined via a Fabry-Perot analysis of the transmission spectra shown in Fig.~\ref{Fig_reflection_transmission}, obtained by THz frequency-domain spectroscopy (THz-FDS). Independent measurements using THz time-domain spectroscopy (THz-TDS) extend to frequencies beyond 1~THz (solid lines). The diamonds depict the dc conductivity determined from van der Pauw measurements. The Corbino spectroscopy data are smoothened using a running average.  The upper abscissa indicates the principal quantum number $n$ of the transition $|n,C\rangle - |n-1,C\rangle$ with transition frequency $f$. Inset: Time-domain spectroscopy data of the imaginary part of the optical conductivity $\sigma_2$.}
	\label{Fig_opticalconductivity}
\end{figure}

Using the same model for the THz data, but now only for narrow ranges around the Fabry-Perot transmission maxima, we can determine the complex optical conductivity $\sigma = \sigma_1 + i \sigma_2$ of the ITO films as a function of frequency. The results are depicted in Fig.~\ref{Fig_opticalconductivity} by the filled circles. In the same plot, we also show the THz conductivity obtained from the THz time-domain spectroscopy (THz-TDS). The latter extends to higher frequencies, and therefore allows us to detect more of the linear-in-frequency $\sigma_2$ increase (see inset to Fig.\ \ref{Fig_opticalconductivity}) which provides confirmation of the above estimate for $\Gamma$. Within the experimental error, the THz $\sigma_1$ of each of the four samples can be considered frequency independent and consistent with the respective dc data, which are additionally shown as diamonds in Fig.~\ref{Fig_opticalconductivity}.

Finally, we include the GHz conductivity data ($f<25 \, \rm{GHz}$) obtained from Corbino spectroscopy in Fig.\ \ref{Fig_opticalconductivity} and find that they properly match the THz and dc results, i.e.\ also the GHz $\sigma_1$ of the investigated ITO films can be considered frequency independent. The GHz spectra also show certain extrinsic features, which are often found in Corbino spectroscopy and which do not affect the significance of the data. Specifically, the strong resonances e.g.\ around 10~GHz stem from three-dimensional standing-wave modes mostly confined within the dielectric substrate \cite{Felger2013}, while the smooth downturn towards lower frequencies for sample B is due to non-perfect Corbino contacts with capacitive contribution \cite{Steinberg2008}. As documented in Table \ref{Tab_ITOsamples}, the much larger thickness of sample D directly leads to substantially lower sheet resistance compared to the other samples, which makes the measurements of this sample much more sensitive to errors of the three-standard Corbino calibration \cite{Scheffler2005}.

These experimental results confirm that our modeling within Drude terms, where $\sigma_1 \approx \sigma_\textrm{dc}$ is essentially frequency independent and $\sigma_2 \ll \sigma_1$, properly describes the entire GHz and THz regimes. Consequently, the computed reflectivity curves in Fig.~\ref{Fig_reflection_transmission}(b) form a reliable basis for evaluating the prospects of ITO thin films to realize long-lived circular Rydberg states.

\section{Prospects for a circular Rydberg atom quantum simulator at room temperature}

The presented GHz and THz data fully characterize the low-energy electrodynamics of ITO that is relevant for the proposed stabilization of high-$n$ circular Rydberg states. The optical conductivity shown in Fig.~\ref{Fig_opticalconductivity} is a material property of ITO, and as such could be used for further studies aiming to relate electronic properties to other material characteristics such as composition or atomic structure. These results can also be directly compared to previous related studies on ITO, in particular using THz time-domain spectroscopy \cite{Yang2013, Wang2015}.

In view of realizing long-lived circular Rydberg states, one has to consider these ITO material properties in the specific geometry of application, which includes the glass substrate and the ITO films with their respective thicknesses. Here, both the explicitly measured THz data as well as our theory model for reflectivity throughout the broader GHz and THz regime, which is fully consistent with the experimentally determined GHz and THz optical conductivity, demonstrate that these optically transparent ITO-coated glass samples indeed strongly reflect at GHz and THz frequencies \cite{Bauer2002}. Note that the finite thickness of the glass substrate induces Fabry-Perot resonances, which have to be considered because of the single-mode nature of Rydberg transitions relevant for the radiative decay. Tailored choice of the glass thickness may thus allow us to tune the GHz reflectivity at the relevant Rydberg transition frequencies. Yet, for the ITO films with highest reflectivity, the modulation due to the substrate becomes rather small. For the largest obtained reflectivity (film D, $R=0.97$ for $f < 100 \, \rm{GHz}$), the predicted lifetime for a circular Rydberg state inside the cavity at room temperature ($T=300 \, \rm{K}$) reaches 6.3~ms for $n=70$ and 17.0~ms for $n=100$ (\textit{cf.} Fig.~\ref{Fig_calculation_lifetime}). For the much thinner film C ($R=0.9$ for $f < 100 \, \rm{GHz}$), the calculated room-temperature lifetime drops to 4.1~ms for $n=70$ and 9.7~ms for $n=100$, which still provides decay rates that are more than an order of magnitude smaller compared to free space. Note that efficient production of very high-$n$ circular states up to $n=70$ has been achieved only recently using crossed-field methods \cite{Morgan2018}, but reaching higher $n$ remains to be demonstrated.

The high transmission of the investigated ITO films in the visible regime allows for combining the stabilization of the circular states with high numerical aperture optical access, which is essential to project tweezer arrays and to collect atomic fluorescence for readout (\textit{cf.} Tab.~\ref{Tab_ITOsamples}). Evidently, trapping of the circular Rydberg atoms is also necessary to fully exploit their long lifetime. Indeed, circular states appear to be well suited for efficient trapping in optical tweezers. First, light-induced ionization is strongly suppressed due to the large orbital angular momentum \cite{Nguyen18,Poirier1988}. Second, the ponderomotive force experienced by the quasi-free Rydberg electron typically results in a repulsive potential and requires hollow-core laser beams for optical trapping \cite{Cortinas19,Barredo2020}. For circular states of sufficiently high $n$, however, the Bohr-like electron orbital can become comparable to the waist of a typical Gaussian-shaped optical tweezer, which first reduces the repulsive potential with increasing $n$ and finally turns it into an attractive one. Exploiting a two-valence electron atom provides even richer means for tailoring the tweezer trap \cite{Wilson2019}.

Finally, let us comment on the expected interaction strength between circular states held in adjacent tweezers. To this end, we compute the pair-interaction for a geometry akin to Fig.~\ref{Fig_decay_paths}(a), i.e. with the orbital plane oriented parallel with the capacitor plates and the two circular Rydberg atoms residing side-by-side \cite{EndnotePairInteraction}. For the pair-state $|n,C ; n,C\rangle$, i.e. when both atoms are in the same manifold, one obtains van-der-Waals interaction shifts of 1.1~MHz (2.1~MHz) at an atomic distance of $r=7 \, \mu \rm{m}$ ($r=12 \, \mu m$) for $n=70$ ($n=100$) exhibiting characteristic $1/r^6$ scaling. For the pair prepared in adjacent manifolds, i.e. $|n,C ; n+1,C\rangle$, and for the same distances we find a dominant dipole-dipole exchange interaction as large as $\sim 14$~MHz ($\sim 17$~MHz) for $n=70$ ($n=100$). The strong interactions may thus allow for reaching up to $\sim 10^4-10^5$ coupling times within the predicted lifetimes in the stabilizing capacitor. Again, we stress that these estimates are obtained for a room temperature environment without cryogenic cooling.

\section{Conclusion and Outlook}

In conclusion, we have elaborated a new approach to stabilize high-$n$ circular Rydberg states against spontaneous and blackbody-induced decay in a transparent, microwave-reflecting cavity formed by ITO-coated glass plates. To this end, we have performed a detailed characterization of the optical properties of ITO thin films in the GHz and THz domain. We have found that high reflectivities at GHz frequencies can be obtained, which allow for reaching circular-state lifetimes up to tens of milliseconds in a room temperature setup, while transparency of the films in the visible spectral range grants vast optical access. We have discussed the prospects of our findings in view of a room-temperature quantum simulator based on circular Rydberg states. While the explicit requirement of high GHz reflectivity investigated here is rather unusual for ITO films, one may build on the existing expertise in materials optimization for transparent conductors, and in particular ITO films, to further increase the dc conductivity combined with high transmission at visible wavelengths \cite{Chopra1983,Hamberg1985,Freeman2000}. Here, a lower scattering rate is a desired but challenging to achieve material property, while the film thickness can be adjusted rather easily to balance GHz reflectivity and visible transmission.

\section*{Acknowledgments}

We are indebted to T. Pfau for generous support, and thank S. Weber and H. P. B\"{u}chler for fruitful and stimulating discussions and G. Untereiner for sample preparation. We acknowledge support from Deutsche Forschungsgemeinschaft [Project No. PF 381/17-1, part of the SPP 1929 (GiRyd)] and the Carl Zeiss Foundation via IQST. F. M. is indebted to the Baden-W\"urttemberg-Stiftung for the financial support by the Eliteprogramm for Postdocs.


\begin{references}
\bibitem{Labuhn2016} H. Labuhn, D. Barredo, S. Ravets, S. de L\'{e}s\'{e}leuc, T. Macr\`{i}, T. Lahaye, and A. Browaeys, Nature \textbf{534}, 667 (2016).
\bibitem{Bernien2017} H. Bernien, S. Schwartz, A. Keesling, H. Levine, A. Omran, H. Pichler, S. Choi, A. S. Zibrov, M. Endres, M. Greiner, and V. Vuleti\'{c}, and M. D. Lukin, Nature \textbf{551}, 579 (2017).
\bibitem{Leseleuc18} S. de L\'{e}s\'{e}leuc, V. Lienhard, P. Scholl, D. Barredo, S. Weber, N. Lang, H. P. B\"{u}chler, T. Lahaye, and A. Browaeys, Science \textbf{365}, 775 (2019).
\bibitem{Levine19} H. Levine, A. Keesling, G. Semeghini, A. Omran, T. T. Wang, S. Ebadi, H. Bernien, M. Greiner, V. Vuleti\'{c}, H. Pichler, and M. D. Lukin, Phys. Rev. Lett. \textbf{123}, 170503 (2019).
\bibitem{Graham19} T. M. Graham, M. Kwon, B. Grinkemeyer, Z. Marra, X. Jiang, M. T. Lichtman, Y. Sun, M. Ebert, and M. Saffman, Phys. Rev. Lett. \textbf{123}, 230501 (2019).
\bibitem{Endres16} M. Endres, H. Bernien, A. Keesling, H. Levine, E. R. Anschuetz, A. Krajenbrink, C. Senko, V. Vuletic, M. Greiner, and M. D. Lukin, Science \textbf{354}, 1024 (2016).
\bibitem{Barredo16} D. Barredo, S. de L\'{e}s\'{e}leuc, V. Lienhard, T. Lahaye, and A. Browaeys, Science \textbf{354}, 1021 (2016).
\bibitem{Barredo18} D. Barredo, V. Lienhard, S. de L\'{e}s\'{e}leuc, T. Lahaye, and A. Browaeys, Nature \textbf{561}, 79 (2018).
\bibitem{Nguyen18} T. L. Nguyen, J. M. Raimond, C. Sayrin, R. Corti\~{n}as, T. Cantat-Moltrecht, F. Assemat, I. Dotsenko, S. Gleyzes, S. Haroche, G. Roux, Th. Jolicoeur, and M. Brune, Phys. Rev. X \textbf{8}, 011032 (2018).
\bibitem{Hulet83} R. Hulet and D. Kleppner, Phys. Rev. Lett. \textbf{51}, 1430 (1983).
\bibitem{Brecha93} R. J. Brecha, G. Raithel, C. Wagner, and H. Walther, Opt. Commun. \textbf{102}, 257 (1993).
\bibitem{Nussenzveig93} P. Nussenzveig, F. Bernardot, M. Brune, J. Hare, J. M. Raimond, S. Haroche, and W. Gawlik, Phys. Rev. A \textbf{48}, 3991 (1993).
\bibitem{HarocheBook06}S. Haroche and J.-M. Raimond, \textit{Exploring the Quantum: Atoms, Cavities, and Photons} (Oxford U. Press, New York, 2006).
\bibitem{Dietsche19} E. K. Dietsche, A. Larrouy, S. Haroche, J. M. Raimond, M. Brune, and S. Gleyzes, Nat. Phys. \textbf{15}, 326 (2019).
\bibitem{Anderson13} D. A. Anderson, A. Schwarzkopf, R. E. Sapiro, and G. Raithel, Phys. Rev. A \textbf{88}, 031401(R) (2013).
\bibitem{Cortinas19} R. G. Corti\~{n}as, M. Favier, B. Ravon, P. M\'{e}haignerie, Y. Machu, J. M. Raimond, C. Sayrin, and M. Brune, Phys. Rev. Lett. \textbf{124}, 123201 (2020).
\bibitem{Cantat20} T. Cantat-Moltrecht, R. Corti\~{n}as, B. Ravon, P. M\'{e}haignerie, S. Haroche, J.-M. Raimond, M. Favier, M. Brune, and C. Sayrin,  Phys. Rev. Research \textbf{2}, 022032(R) (2020).
\bibitem{Kleppner85} D. Kleppner, Phys. Rev. Lett. \textbf{47}, 233 (1981).
\bibitem{Hulet85} R. G. Hulet, E. S. Hilfer, and D. Kleppner, Phys. Rev. Lett. \textbf{55}, 2137 (1985).
\bibitem{HartnagelBook1995} H. L. Hartnagel, A. L. Dawar, A. K. Jain, and C. Jagadish, \textit{Semiconducting Transparent Thin Films} (IOP Publishing, London, UK, 1995). 
\bibitem{Tahar1998} R. B. H. Tahar, T. Ban, Y. Ohya, and Y. Takahashi, J. Appl. Phys. \textbf{83}, 2631 (1998).
\bibitem{GallagherBook94} T. F. Gallagher, \textit{Rydberg Atoms} (Cambridge University Press, Cambridge, England, 1994). 
\bibitem{Haroche92} S. Haroche, in \textit{Fundamental Systems in Quantum Optics, Les Houches Summer School, Session LIII}, J. Dalibard, J.-M. Raimond, and J. Zinn-Justin, eds. (North Holland, Amsterdam, 1992).
\bibitem{Hinds94} E. A. Hinds, in \textit{Advances in Atomic and Molecular Physics, Supplement 2}, P. Berman ed. (Academic Press, New York, 1994).
\bibitem{Mozley05} J. Mozley, P. Hyafil, G. Nogues, M. Brune, J.-M. Raimond, and S. Haroche, Eur. Phys. J. D \textbf{35}, 43 (2005).
\bibitem{Freeman2000} A. J. Freeman, K. R. Poeppelmeier, T. O. Mason, R. P. H. Chang, and T. J. Marks, MRS Bulletin \textbf{25}(8), 45 (2000).
\bibitem{Gordon2000} R. G. Gordon, MRS Bulletin \textbf{25}(8), 52 (2000).
\bibitem{Minami2005} T. Minami, Semicond. Sci. Technol. \textbf{20}, S35 (2005).
\bibitem{Takizawa1999} K. Takizawa and O. Hashimoto, 	IEEE Trans. Microw. Theory Tech. \textbf{47}, 1137 (1999).
\bibitem{Bauer2002} T. Bauer, J. S. Kolb, T. L\"offler, E. Mohler, H. G. Roskos, and U. C. Pernisz, J. Appl. Phys. \textbf{92}, 2210 (2002).
\bibitem{Yang2013} C.-S. Yang, M.-H. Lin, C.-H. Chang, P. Yu, J.-M. Shieh, C.-H. Shen, O. Wada, and C.-L. Pan, IEEE J. Quantum Electron. \textbf{49}, 677 (2013).
\bibitem{Brown2015} E. R. Brown, W-D. Zhang, H. Chen, and G. T. Mearini, Appl. Phys. Lett. \textbf{107}, 091102 (2015).
\bibitem{Wang2015} T. Wang, M. Zalkovskij, K. Iwaszczuk, A. V. Lavrinenko, G. V. Naik, J. Kim, A. Boltasseva, and P. U. Jepsen, Opt. Mater. Express \textbf{5}, 566 (2015).
\bibitem{Sheokand2017} H. Sheokand, S. Ghosh, G. Singh, M. Saikia, K. V. Srivastava, J.
Ramkumar, and S. A. Ramakrishna, J. Appl. Phys. \textbf{122}, 105105 (2017).
\bibitem{Zhang2017} C. Zhang, Q. Cheng, J. Yang, J. Zhao, and T. J. Cui, Appl. Phys. Lett. \textbf{110}, 143511 (2017).
\bibitem{Lai2017} S. Lai, Y. Wu, X. Zhu, W. Gu, and W. Wu, IEEE Photonics J. \textbf{9}, 5503310 (2017).
\bibitem{Hamberg1984} I. Hamberg, C. G. Granqvist, K.-F. Berggren, B.E. Sernelius, and L. Engstr\"om, Phys. Rev. B \textbf{30}, 3240 (1984).
\bibitem{Gorshunov2005} B. Gorshunov, A. Volkov, I. Spektor, A. Prokhorov, A. Mukhin, M. Dressel, S. Uchida, and A. Loidl, Int. J. Infrared Millim. Waves \textbf{26}, 1217 (2005).
\bibitem{Pracht2013} U. S. Pracht, E. Heintze, C. Clauss, D. Hafner, R. Bek, S. Gelhorn, D. Werner, M. Scheffler, M. Dressel, D. Sherman, B. Gorshunov, K. S. Il'in, D. Henrich, M. Siegel, IEEE Trans. THz Sci. Technol. \textbf{3}, 269 (2013).
\bibitem{Hangyo2005} M. Hangyo, M. Tani, and T. Nagashima, Int. J. Infrared Millim. Waves \textbf{26}, 1661 (2005).
\bibitem{Scheffler2005} M. Scheffler and M. Dressel, Rev. Sci. Instrum. \textbf{76}, 074702 (2005).
\bibitem{EndnoteVacuLayer} The samples were obtained from VacuLayer Corp.
\bibitem{Felger2013} M. M. Felger, M. Dressel, and M. Scheffler, Rev. Sci. Instrum. \textbf{84}, 114703 (2013).
\bibitem{Steinberg2008} K. Steinberg, M. Scheffler, and M. Dressel, Phys. Rev. B \textbf{77}, 214517 (2008).
\bibitem{Morgan2018} A. A. Morgan, V. Zhelyazkova, and S. D. Hogan, Phys. Rev. A \textbf{98}, 043416 (2018).
\bibitem{Poirier1988} M. Poirier, Phys. Rev. A \textbf{38}, 3484 (1988).
\bibitem{Barredo2020} D. Barredo, V. Lienhard, P. Scholl, S. de L\'{e}s\'{e}leuc, T. Boulier, A. Browaeys, and T. Lahaye, Phys. Rev. Lett. 124, 023201 (2020).
\bibitem{Wilson2019} J. Wilson, S. Saskin, Y. Meng, S. Ma, R. Dilip, A. Burgers, and J. Thompson, arXiv:1912.08754 (2019).
\bibitem{EndnotePairInteraction} The pair-potentials are computed in the presence of a small stabilizing electric field $E=20 \, \rm{mV/cm}$ and a magnetic field $B=10 \, \rm{G}$, which splits the degeneracy with elliptical states and thereby suppresses interaction-induced mixing \cite{Nguyen18}. Both fields point orthogonal to the capacitor plates.
\bibitem{Chopra1983} K. L. Chopra, S. Major, and D. K. Pandya, Thin Solid Films \textbf{102}, 1 (1983).
\bibitem{Hamberg1985} I. Hamberg and C. G. Granqvist, Appl. Opt. \textbf{24}, 1815 (1985).

\end{references}
\end{document}